\documentclass[11pt,twoside]{article}

\usepackage{asp2006_m}
\usepackage{epsf}
\usepackage{psfig}
\usepackage{lscape}

\begin{document}
\title{3D spectroscopy of the ionized gas kinematics in galactic rings.}   
\author{Alexei Moiseev}   
\affil{Special Astrophysical Observatory, Nizhnij Arkhyz 369169, Russia
}    

\begin{abstract} 
The kinematics of galactic rings were studied with a scanning Fabry-Perot
interferometer   mounted in the multi-mode focal reducer  SCORPIO \citep{moiseev:afan}
at  the SAO RAS 6-m telescope. The analysis of the ionized gas velocity fields allows us
to understand the nature of the ring formation in several galaxies. The  different types
of the rings in  the presented objects  (resonanced, collisional, polar)  were caused by
the various sorts  of interactions: merging, head-on collisions.
\end{abstract}



\section{NGC 7742 \citep{moiseev:sil}}

The global gaseous disk in Sb  galaxy NGC 7742 counter-rotates relative to  the stellar
one; the galaxy possess two exponential stellar disks with different scalelengths. We
suggest  that a past minor  merger is the probable cause for the star-forming ring
without global bar; the ring might be produced as a resonance feature by tidally induced
oval distortions of the global stellar disks. \citet{moiseev:knapen} also offered  the
similar mechanism for the ring in the galaxy NGC~278. This idea agrees with the results
of numerical simulations of the density re-distribution during the galactic tidally
interaction, if the satellite moves in the plane of main galaxy but with reverse
direction of the rotation \citep{moiseev:tutukov}.

\section{Arp 10 \citep*{moiseev:biz}}

 The  pecular galaxy Arp~10  has two rings (the inner and outer one), and extended outer
arc.  The H$\alpha$ velocity field  shows evidence for significant radial motions in
both outer and inner galactic rings. We fit a model velocity field taking into account
the regular rotation and projection effects.  The expansion velocity of the NW part of
the outer ring exceeds 100\,km\,s$^{-1}$, whereas it attain only 30\,km\,s$^{-1}$ at the
SE part. This asymmetric   may be a result a  substantially  off-center collision with
the intruder. We identified the nucleus of a companion, which  was an early-type spiral
galaxy before the collision with mass about $1/4$ the mass of target galaxy. The sizes
of the inner and outer rings, maximum expansion velocity of the outer ring, and radial
profile of the gas circular velocity can be reproduced by a near-central collision with
the intruder galaxy, which occurred approximately 85 Myr ago. The companion passed
through the disk of Arp10 about 3 kpc from the nucleus and generated an expanding
asymmetric density wave in the disk of Arp10.

\section{Arp 212 \citep{moiseev:mois}}

\begin{figure}
\plotone{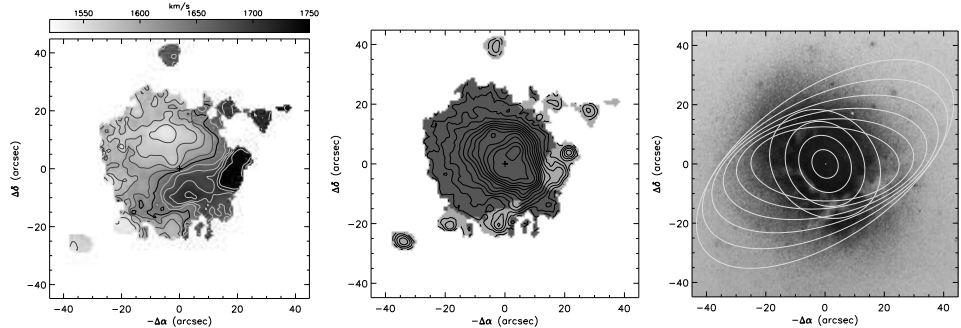} \caption{Arp 212: the H$\alpha$  velocity field (left); the mask
of kinematic  distinct components with H$\alpha$ isophotes (middle);   the layout of the
circular orbits orientation is  superposed  on the photographic from Arp's atlas
(right).}
\label{mois:fig1}       
\end{figure}

Sa/S pec galaxy Arp 212 happens to have a great deal of gas and dust associated with
violent star formation. H$\alpha$ emission concentrates in the center region and also in
individual knots distributed in the external curvilinear structures which are possible
signs of recent interaction \citep{moiseev:cairos}. On the ionized gas velocity field we
detect regular rotation at $r<3$ kpc. The line-of-sight velocities in the external knots
have systematic (up to 70-80\,km\,s$^{-1}$) deviations from the inner part rotation
pattern. The analysis in the frame of tilted-ring approximation  shows that external
knots rotate on stable orbits, the systemic velocity and   amplitude of the rotation
curve agree with same parameters on the inner radii. However, the  spatial orientation
of the rotation plane differs for inner   and outer  parts. The angle between the `main
plane' and the `outer ring' reaches  $\Delta i>75^\circ$ at the distances $r=4-5$ kpc.
Therefore the  outer HII knots in the Arp 212  have an external origin and rotate  in a
distinct ring. The inner parts of this ring are warped, the outer ones may rotate in the
plane nearly orthogonal to the plane of the inner gaseous disk (see
Fig.~\ref{mois:fig1}).

\acknowledgements 
This work was partly supported by the Russian  Foundation for Basic Research (project  06-02-16825) and by the
grant of President of Russian Feredration (MK1310.2007.2)

\end{document}